\documentclass[superscriptaddress,preprint,nofootinbib,nobibnotes,eprint,amsmath,amssymb,aps,prb,citeautoscript,titlepage,nobalancelastpage,floats,final,eqsecnum]{revtex4-2}

\usepackage[utf8]{inputenc}

\usepackage{amsmath}
\usepackage{amssymb}
\usepackage{amsfonts}

\usepackage{latexsym}
\usepackage{slashed}
\usepackage[svgnames]{xcolor}
\usepackage{soul}
\usepackage{braket}
\usepackage{MnSymbol}
\usepackage{wasysym}
\usepackage{mathrsfs}  
\usepackage{physics}
\usepackage{enumitem}      
\usepackage[title]{appendix}
\usepackage{tikz}

\usepackage{graphics}
\graphicspath{ {graphs/} }

\usepackage{microtype}
\usepackage{multirow}

\usepackage{hyperref}
\usepackage{natbib}

\hypersetup{
	pdfencoding=auto,
	psdextra,
	colorlinks=true,
	linkcolor=Blue,
	citecolor=DarkRed,
	urlcolor=Blue
}

\usepackage{mathtools}

\begin{document}
	
	\title{\Large{{\sc Lattice Realization of Twist Defects in a $\mathbb{Z}_2\times \mathbb{Z}_2$ Topological Order }}}
	
	\author{Gustavo M. Yoshitome}
	\email{gustavoyoshitome@gmail.com }
	\affiliation{Departamento de Física, Universidade Estadual de Londrina, Londrina, PR, Brasil}

	\begin{abstract}	
		
In this work, we explore a microscopic realization of three types of anyonic symmetries in a $\mathbb{Z}_2\times\mathbb{Z}_2$ topological order, corresponding to a double toric code. These symmetries act as nontrivial permutations on the anyon labels of the parent state. We consider a setup consisting of two decoupled Wen plaquette models stacked on top of each other and introduce dislocations that modify the Hamiltonian, giving rise to localized twist defects, eventually inducing interactions between the layers. In this context, branch cuts act as sources of anyon permutations when they cross it. We characterize the defects by calculating their quantum dimensions, and we also consider double loop operators around them that allow us to determine the non-Abelian fusion rules between the defects, including when they carry different anyon permutations.

	\end{abstract}

	\maketitle

	\tableofcontents

\section{Introduction}

Topological order  is a phase of matter characterized by a topological ground state degeneracy, which is robust against any perturbation, making it a viable candidate for storing quantum information \cite{Dennis_2002,preskill2012quantumcomputingentanglementfrontier,Fujii:2015wia,Wen_2017}.

Another fascinating aspect of this phase of matter is that it supports quasiparticle excitations called anyons, whose fusion and braiding properties go beyond those of conventional particles \cite{PhysRevB.40.7387,Kitaev:1997wr}. In addition to these intrinsic excitations, topological orders may also possess anyonic symmetries, which are transformations that permute anyon types while preserving their fusion and braiding structure \cite{Khan_2014,Barkeshli:2014cna,Khan:2016fth,Teo:2015xla}. When such symmetries are implemented, they give rise to twist defects, extrinsic objects that act as sources of branch cuts across which anyons are permuted. These defects exhibit rich and often non-Abelian zero modes, even when the underlying topological order is Abelian \cite{Teo_2016,Barkeshli:2012pr,Barkeshli_2012}, and have been the subject of extensive study due to their conceptual importance and potential applications in topological quantum computation \cite{Iulianelli:2024icl,Nayak_2008,Field:2018qtg}.

A particularly relevant model that captures the essence of topological order is the toric code, due to its simple anyonic structure.
It has an anyonic symmetry which can be implemented via twist defects on the lattice by considering dislocations on a Wen plaquette model. In this context, it is possible to obtain non-Abelian fusion rules, quantum dimension, and topological entanglement entropy directly from the lattice model \cite{You:2012sfg,You:2012wz,Bombin:2010xn,Brown:2013goa,Kitaev_2012,yan2024generalizedkitaevspinliquid}. In the realm of microscopic models, Kitaev's honeycomb model \cite{Kitaev:2005hzj,Kitaev:2009war} has also been studied in the presence of dislocations, where Majorana zero modes emerge as consequence of the defects \cite{Petrova:2013caa}. Recent works have also explored junctions between toric codes to study anyonic permeability in non-commuting defects \cite{Bhattacharjee:2025tak}. And in other point of view, they can be studied as emergent higher symmetries \cite{Barkeshli:2022wuz,Barkeshli:2022edm,Barkeshli:2023bta}.

Outside of microscopic realizations, twist defects have also been implemented in setups involving quantum Hall states and superconductors. In this case, effective descriptions are capable of characterizing the static non-Abelian modes localized at the core of defects. Therefore, they  could potentially provide realizations of non-Abelian zero modes\cite{Vaezi_2013,Clarke_2013,Cao:2023tse,Cheng_2012,Katzir:2020trf,Lindner_2012,Yoshitome:2025cit}.

In this work, we construct dislocations on the lattice of a $\mathbb{Z}_2 \times \mathbb{Z}_2$ topological order, that reproduce three types of anyonic symmetries. We consider a bilayer construction of two Wen plaquette models, that realizes a double toric code topological order, which has a rich set of anyonic symmetries. By introducing branch cuts, we are capable of modifying the exactly solvable Hamiltonian into one that allows for the presence of localized defects on the lattice.

We classify each defect type from the permutation they employ on the parent state, and classify their properties by using operators defined by the microscopic Hamiltonian. Bilayer systems become particularly interesting in the presence of defects, since they can introduce nontrivial interactions between the layers \cite{Teo_2014,May_Mann_2020}. We also obtain the fusion rules for the static non-Abelian modes that are trapped at the core of the defects by using the algebra of loop operators, which are capable of characterizing all excitations in the spectrum. In particular, we compute the fusion rule between defects carrying different types of anyon permutations, which reveals a projective representation of the symmetry actions due to an Abelian anyon associated with the trivalent junction of the fusion process.

Previous studies have classified twist defects in the single layer Wen plaquette model, which realizes a $\mathbb{Z}_2$ topological order. In contrast, the bilayer Wen plaquette model considered in this work realizes a $\mathbb{Z}_2 \times \mathbb{Z}_2$ topological order and can support multiple distinct types of twist defects within a single lattice realization, whereas the single layer model hosts only one. Although twist defects in $\mathbb{Z}_2 \times \mathbb{Z}_2$ topological orders have previously been studied from an abstract perspective \cite{Kitaev_2012,Moradi:2022lqp} and in specific lattice realizations such as the color code \cite{Kesselring:2018jbv}, our work provides a full characterization of three types of defects and their fusion rules on the bilayer Wen plaquette mode.

\section{Dislocations on  Wen Plaquette Model \label{sec3}}
	
Twist defects can be realized in any system with nontrivial anyon content, provided it is sufficiently rich to support anyonic symmetries. An example of a topological order exhibiting such defects is the toric code. We now review a simple lattice model, the Wen plaquette model \cite{Wen:2003yv}, that realizes twist defects through dislocations.

The Wen plaquette model is defined on a square lattice and realizes a $\mathbb{Z}_2$ topological order. Its Hamiltonian is given by
	\begin{equation}
		H  = -J\,\, \sum_{\vec{r}} F_{\vec{r}} = -J\,\, \sum_{\vec{r}} \begin{tikzpicture}[baseline={(current bounding box.center)}]

			\draw (0,0) -- (1.3,0) -- (1.3,1.3) -- (0,1.3) -- cycle;

			\node[below left, xshift=6.0pt, yshift=-0.2pt] at (0,0) {$X_{\vec{r}}$};
			\node[below right, xshift=-6.0pt, yshift=-0.2pt] at (1.3,0) {$Z$};
			\node[above right, xshift=-6.0pt, yshift=0.2pt] at (1.3,1.3) {$X$};
			\node[above left, xshift=6.0pt, yshift=0.2pt] at (0,1.3) {$Z$};
			
		\end{tikzpicture}.
\end{equation}
where $J$ is a positive constant and $F_{\vec{r}} = X_{\vec{r}} Z_{\vec{r}+\hat{y}} X_{\vec{r}+\hat{x}+\hat{y}} Z_{\vec{r}+\hat{x}}$.

On an even $\times$ even lattice, the system admits a checkerboard bipartition into two sublattices. This structure naturally defines two distinct anyons, $e$ and $m$, which reside on different sublattices. The full anyon content is $\{1, e, m, \psi\}$, where $\psi = e \times m$ and $e\times e = m\times m = 1$.

We define the parity of a sublattice by the parity of its sites \footnote{The parity of a site $\vec{r}$ is defined by the parity of the sum of its coordinates.}, and the parity of a plaquette as the parity of its bottom left site. Anyonic excitations can be created through string operators, called Wilson lines, defined on either the even plaquettes or the odd plaquettes. Explicitly, they read
\begin{equation}
    W_e = \prod_{\vec{r} \in \gamma_{\mathrm{even}}} \begin{cases} Z_{\vec{r}} & \text{if $\vec{r}$ is even} \\
X_{\vec{r}} & \text{if $\vec{r}$ is odd} \end{cases}
\label{W}
\end{equation}
\begin{equation}
\label{Wm}
    W_m = \prod_{\vec{r} \in \gamma_{\mathrm{odd}}} \begin{cases} Z_{\vec{r}} & \text{if $\vec{r}$ is odd} \\
X_{\vec{r}} & \text{if $\vec{r}$ is even} \end{cases},
\end{equation}
where $\gamma_{\mathrm{even}}$ and $\gamma_{\textrm{odd}}$ are paths defined on even and odd plaquettes, respectively.

The operator on Eq. (\ref{W}) creates $e$ excitations at its endpoints on the even sublattice, and the operator in Eq. (\ref{Wm}) creates $m$ excitations at its endpoints in the odd sublattice.

In order to move an excitation to another site, one concatenates an additional Wilson line that creates the same type of excitation, with endpoints located at the position of the existing excitation and at the target site. This process annihilates the excitation at one endpoint and creates an identical excitation at the other.

If any of the string operators in Eqs. (\ref{W}) and (\ref{Wm}) are defined on closed loops, they are instead referred to as Wilson loops, denoted by $\Gamma_e$ and $\Gamma_m$, respectively.

This model possesses a well known anyonic symmetry, the electric-magnetic duality, which exchanges the anyon labels $e$ and $m$ \cite{Buerschaper:2010yf,Vidal:2009ma,Kitaev_2012,Teo:2015xla}, and we denote this symmetry by $\sigma$. It gives rise to two distinct types of twist defects, which can be implemented as lattice dislocations.

\begin{figure}[h]
	\centering
	\includegraphics[width=0.45\linewidth]{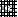}
	\caption{A dislocation defect localized at the pentagon operator $P_{\vec{r}}^{\pm}$. When an $e$ quasiparticle crosses the cut lines, it mutates into an $m$ quasiparticle.}
	\label{fig:dislocation}
\end{figure}

Consider the lattice in the presence of such a dislocation, as shown in Fig.~\ref{fig:dislocation}. This defect has a trivalent endpoint corresponding to the pentagon operator
\begin{equation}
\label{2.4}
	P^{\pm} = \pm Y_1X_2Z_3X_4Z_5 = \pm\begin{tikzpicture}[baseline={(current bounding box.center)}]

	\def\L{1.3}

	\coordinate (A) at (0,0);
	\coordinate (B) at (\L,0);
	\coordinate (C) at (\L,\L);
	\coordinate (D) at (0,\L);
	\coordinate (E) at (-0.5*\L,0.5*\L);  
    
	\draw (A) -- (B) -- (C) -- (D) -- (E) -- cycle;

	\node[below left, xshift=6.0pt, yshift=-1.5pt] at (A) {$X_2$};
	\node[below right, xshift=-6.0pt, yshift=-1.5pt] at (B) {$Z_4$};
	\node[above right, xshift=-6.0pt, yshift=1.5pt] at (C) {$X_5$};
	\node[above left, xshift=6.0pt, yshift=1.5pt] at (D) {$Z_3$};
	\node[left, xshift=-1.5pt] at (E) {$Y_1$};
	
	\end{tikzpicture}.
\end{equation}

The two possible signs of the pentagon operator correspond to distinct defect types. These can be distinguished by braiding an anyon around a single defect. For example, consider transporting an $e$ particle around the defect using a Wilson line operator $\hat{\Theta}$. Due to the action of the anyonic symmetry, the particle must encircle the defect twice in order to return to its original type and annihilate.

This process generates a self intersection of the Wilson line, corresponding to the operator $ZX = iY$. Since the Wilson loop can be expressed as the product of all plaquette operators within the enclosed region, it follows that $\hat{\Theta} \ket{\text{GS}} = i \ket{\text{GS}}$ for $P^+$, and $\hat{\Theta} \ket{\text{GS}} = -i \ket{\text{GS}}$ for $P^-$.

\subsection{\label{secA} Fusion Rules of the Twisted Wen Model}

For the anyonic symmetry $\sigma$, there are two types of defects, which we denote by $\sigma_0$ and $\sigma_1$ \cite{You:2012sfg,Teo:2015xla}. These defect types differ by an anyon charge in the parent state that they carry. Therefore, they can be seen as a bound state of a defect with an anyon. We label them $\sigma_0$ that carries no charge, while $\sigma_1$ is bound to an $e$ anyon. 

The quantum dimension of $\sigma_i$, $d_{\sigma_i}$, can be calculated by evaluating the asymptotic growth of the ground state degeneracy as defects, i.e. the dislocations introduced by the pentagon operator in Eq. (\ref{2.4}), are introduced into the system.\footnote{Specifically, the quantum dimension of a pair of twist defects is given by $\lim_{n\rightarrow\infty}\left(GSD_n(\sigma_i)\right)^{\frac{1}{n}}$, where $GSD_n(\sigma_i)$ is the ground state degeneracy in the presence of $n$ pairs of $\sigma_i$ defects.} \cite{PhysRevB.88.235103,You:2012sfg} In the absence of dislocations and under periodic boundary conditions, the Wen plaquette model has a ground state degeneracy of $4$ (on lattices with even linear dimensions) \cite{Wen:2003yv}. When a pair of pentagon operators is introduced, one spin becomes unconstrained by any stabilizer. To see this, we associate each plaquette operator with, for example, its bottom-left spin, and similarly associate each pentagon operator with its bottom-left spin. The spin located at $Y_1$ in Eq.~(\ref{2.4}) is not associated with any stabilizer. Consequently, since the Hamiltonian is a sum of stabilizer and pentagon operators, one degree of freedom remains unconstrained, leading to an additional two-fold degeneracy. More generally, introducing $N$ pairs of dislocations increases the ground state degeneracy by a factor of $2^N$. Therefore, the quantum dimension of each twist defects $\sigma_i$ is $d_{\sigma_i}=\sqrt{2}$.

However, note that only one pair of dislocations is necessary to infer the quantum dimension, since the presence of additional defects changes the ground state by the same factor.

The full set of quasiparticles $\{1, e, m, \psi, \sigma_0, \sigma_1\}$ obeys the fusion rules
\begin{equation}
\sigma_i \times e = \sigma_i \times m = \sigma_{i+1},
\end{equation}
\begin{equation}
 \sigma_i \times \psi = \sigma_i,
 \end{equation} 
\begin{equation}
	\sigma_i \times \sigma_i = 1+\psi
\end{equation}

To derive the last fusion rule, note that the total topological charge contained within a region can be determined by evaluating Wilson loop operators $\Gamma_e$ and $\Gamma_m$, corresponding to the worldlines of $e$ and $m$ particles encircling the region.

\begin{figure}
	\centering
	\includegraphics[width=0.4\linewidth]{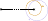}
	\caption{Closed loop of an anyon $e$ or $m$ around the defect. Once it is acted by the defect, the line changes color, corresponding to the change of anyon type.}
	\label{fig:fig1}
\end{figure}

As discussed in the previous section, a Wilson loop of an $e$ or $m$ particle must wind twice around a single defect; otherwise, the process does not return to the ground state subspace, since the particle does not annihilate itself at the end of the trajectory. This process is illustrated in Fig.~\ref{fig:fig1}.

As we have seen, when the operator $\hat{\Theta}$ acts on the ground state, it produces an eigenvalue $i$. We now bring two such defects close together and evaluate the Wilson loop operators $\Gamma_e$ and $\Gamma_m$ around the pair.

\begin{figure}
	\centering
	\includegraphics[width=0.8\linewidth]{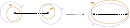}
	\caption{Two paths around each defect can be deformed to a pair of paths linked around both defects.}
	\label{fig:fig2}
\end{figure}

To proceed, consider a path encircling each defect individually. These paths can be smoothly deformed into a pair of linked loops surrounding both defects, one corresponding to the worldline of an $e$ quasiparticle and the other to that of an $m$ quasiparticle. This configuration is depicted in Fig.~\ref{fig:fig2}.

 Unlinking these loops introduces a minus sign due to their mutual statistics, yielding
\begin{equation}
	\label{3.6}
	\Gamma_e \Gamma_m \ket{\text{GS}} = - \hat{\Theta}_1 \hat{\Theta}_2 \ket{\text{GS}}= \ket{\text{GS}}
\end{equation}
Therefore, the total topological charge of the pair can only lie in the fusion channels $1$ or $\psi$.  Since $\sigma_i$ has quantum dimension $\sqrt{2}$, the total quantum dimension of the fusion outcomes in $\sigma_i \times \sigma_i$ must be $2$. Moreover, consistency with $\sigma_i \times \psi = \sigma_i$ rules out the possibilities of each channels $1$ or $\psi$ appearing with multiplicity two.\footnote{Suppose, for instance, that $\sigma_i \times \sigma_i = 2 \psi$. Fusing both sides with $\psi$ would then lead to the fusion rule $\sigma_i\times\sigma_i = 1+1$, which is inconsistent with the original one.} Therefore, we conclude that $\sigma_i \times \sigma_i = 1 + \psi$.

We emphasize that the loop algebra of Wilson operators is sufficient to fully determine the topological charges and fusion channels. This perspective will be employed throughout the remainder of this work to analyze more intricate defect structures.
\begin{equation}
	\text{FUSION CHANNELS} 	\Longleftrightarrow \text{LOOP ALGEBRA}
\end{equation}

\section{\label{secIII}$\mathbb{Z}_2 \times \mathbb{Z}_2$ Topological Order}

We now turn to the description of defects in a $\mathbb{Z}_2 \times \mathbb{Z}_2$ topological order. This phase can be understood as two copies of the toric code, $\mathcal{A}_1 = \{1, e_1, m_1, \psi_1\}$ and $\mathcal{A}_2 = \{1, e_2, m_2, \psi_2\}$ with 
\begin{equation}
	e_1 \times m_1 =\psi_1 \quad \text{and} \quad e_2 \times m_2 = \psi_2.
\end{equation}
The full set of anyons is therefore generated by
\begin{equation}
\mathcal{A} = \mathcal{A}_1 \times \mathcal{A}_2=	\langle e_1, m_1, e_2, m_2 \rangle.
\end{equation}

Fusion and braiding properties are inherited independently from each copy. It is convenient to label the anyons by vectors in the lattice $\mathbb{Z}_2^4$, where fusion corresponds to vector addition. In this representation, a generic excitation is described by a vector $(a_1, a_2, a_3, a_4)$, with $a_i \in \mathbb{Z}_2$. The components correspond, respectively, to the presence of $e_1$, $m_1$, $e_2$, and $m_2$ charges.

This system can be viewed as two layers of the toric code stacked on top of each other. At first glance, the layers appear completely decoupled, as excitations from different layers have trivial mutual braiding. However, this perspective is incomplete, as the full structure of the theory includes nontrivial anyonic symmetries that act as automorphisms of the anyon model. In particular, these symmetries can exchange the two layers or, more generally, mix their anyon labels in a nontrivial way. As we will see, introducing defects associated with such symmetries leads to new sectors where the layer degree of freedom is no longer conserved, giving rise to qualitatively richer fusion and braiding structures.

\subsection{Anyonic Symmetries and Defects}

The set of anyons in a topological order admits a special class of symmetries known as anyonic symmetries. These are permutations of anyon labels that preserve both fusion and braiding properties. The $\mathbb{Z}_2 \times \mathbb{Z}_2$ topological order possesses a large group of such symmetries. Concretely, the full set of symmetries is given by $S_3 \times S_3 \rtimes\mathbb{Z}_2$, where $S_3$ is the group of permutations of three symbols. Therefore, the number of anyonic symmetries is $\vert S_3 \times S_3 \rtimes \mathbb{Z}_2\vert = 72$ \cite{Moradi:2022lqp,Etingof2010FusionHomotopy,NIKSHYCH2014191}.
Here, we focus on a subset of three  symmetries that act nontrivially on anyon species. They are defined by
\begin{align}
	\sigma:\,\, \, \, &(e_1,e_2) \rightarrow (m_1,m_2)\\
	&(m_1,m_2) \rightarrow (e_1,e_2)
\end{align}
\begin{align}
	\rho:\,\, \, \, &(e_1,m_1) \rightarrow (e_2,m_2)\\
	&(e_2,m_2) \rightarrow (e_1,m_1)
\end{align}
\begin{align}
	\xi:\,\, \, \, &(e_1,m_1) \rightarrow (m_2,e_2)\\
	&(m_2,e_2) \rightarrow (e_1,m_1)
\end{align}

All three symmetries are twofold, meaning that applying them twice lead to an identity permutation. The symmetry $\sigma$ corresponds to electric-magnetic duality acting independently on each layer, $\rho$ exchanges the two layers, and $\xi$ combines layer exchange with electric-magnetic duality.

Anyonic symmetries can be physically realized as twist defects. When an anyon crosses such a defect, its type is transformed according to the corresponding permutation. Each symmetry admits multiple defect types, which are classified by equivalence classes of anyons under the symmetry action. These classes can be identified by considering anyon modulo the subgroup generated by $(1 - \lambda)\mathcal{A}$, where $\lambda$ denotes the symmetry action over the anyons. For the symmetries considered in this work, it is precisely the subgroup of invariant anyons under the symmetry. We explain this in detail in Appendix \ref{apA}.

For the symmetry $\sigma$, the defect types are labeled by
\begin{equation}
	\frac{\mathcal{A}}{(1-\sigma)\mathcal{A}} = \{\overline{1},\overline{e_1 } ,\overline{e_2},\overline{e_1e_2}\},
\label{4.9}
\end{equation}
where each element is a coset of anyons identified by the respective symmetry. Thus, there are four defect species, with charge labels
\begin{equation}
  \sigma_0 \sim  (0,0,0,0),\quad \sigma_1 \sim (1,1,0,0),\quad \sigma_2 \sim (0,0,1,1),\quad \sigma_3 \sim(1,1,1,1)
\end{equation}
The first corresponds to a defect associated with the invariant anyon $1$, while the others are obtained by fusing this defect with $e_1$, $e_2$, and $e_1 e_2$, respectively.

For the symmetry $\rho$, we have
\begin{equation}
	\label{4.10}
	\frac{\mathcal{A}}{(1-\rho)\mathcal{A}} = \{\overline{1},\overline{e_1},\overline{ m_1},\overline{ \psi_1}\},
\end{equation} 
corresponding to the charge assignments
\begin{equation}
 \rho_0 \sim (0,0,0,0),\quad \rho_1 \sim(1,0,1,0),\quad \rho_2 \sim (0,1,0,1),\quad \rho_3 \sim (1,1,1,1)
\end{equation}
As before, these arise from a neutral defect associated with the invariant anyon $1$, together with its fusion with $e_1$, $m_1$, and $\psi_1$. 

Similarly, for the symmetry $\xi$, the defect types are labeled by
\begin{equation}
	\label{4.11}
	\frac{\mathcal{A}}{(1-\xi)\mathcal{A}} = \{\overline{1},\overline{e_1} ,\overline{m_1},\overline{\psi_1}\},
\end{equation} 
that corresponds to the charges $\xi_0 \sim (0,0,0,0)$, $\xi_1 \sim(1,0,0,1)$, $\xi_2 \sim (0,1,1,0)$, and $\xi_3 \sim(1,1,1,1)$ under the original anyons. 

Defects carrying trivial topological charge are referred to as bare defects, while all others can be obtained by attaching an anyon string to a bare defect, thereby endowing it with charge.

When a parent topological state is enriched by defects associated with a symmetry $\lambda$, the resulting structure is described by a defect fusion category of the form $\mathcal{C}_1 \oplus \mathcal{C}_{\lambda}$. An important feature of the symmetries $\sigma$, $\rho$, and $\xi$ is that they are closed under composition. Consequently, the full theory involves multiple sectors, and fusion rules are constrained to map objects between the categories $\mathcal{C}_1$, $\mathcal{C}_\sigma$, $\mathcal{C}_\rho$, and $\mathcal{C}_\xi$ \cite{Teo:2015xla}.

\section{Realization of Defects on the Lattice}

To implement these defects on the lattice, we consider a double Wen plaquette model, consisting of two copies stacked on top of each other. On a square lattice, we assign two degrees of freedom, corresponding to qubits, to each vertex. Thus, each site carries a Hilbert space of the form $\mathcal{H}_1 \otimes \mathcal{H}_2$, where each factor is two-dimensional.

The Hamiltonian is given by
\begin{equation}
	H= \sum_{\vec{r} \in \,\text{lattice}} \begin{tikzpicture}[baseline={(current bounding box.center)}]

	\draw (0,0) -- (1.3,0) -- (1.3,1.3) -- (0,1.3) -- cycle;

	\node[below left, xshift=6.0pt, yshift=-0.2pt] at (0,0) {$X^1_{\vec{r}}$};
	\node[below right, xshift=-6.0pt, yshift=-0.2pt] at (1.3,0) {$Z^1$};
	\node[above right, xshift=-6.0pt, yshift=0.2pt] at (1.3,1.3) {$X^1$};
	\node[above left, xshift=6.0pt, yshift=0.2pt] at (0,1.3) {$Z^1$};
	
	\end{tikzpicture}
	+ \begin{tikzpicture}[baseline={(current bounding box.center)}]

	\draw (0,0) -- (1.3,0) -- (1.3,1.3) -- (0,1.3) -- cycle;

	\node[below left, xshift=6.0pt, yshift=-0.2pt] at (0,0) {$X^2_{\vec{r}}$};
	\node[below right, xshift=-6.0pt, yshift=-0.2pt] at (1.3,0) {$Z^2$};
	\node[above right, xshift=-6.0pt, yshift=0.2pt] at (1.3,1.3) {$X^2$};
	\node[above left, xshift=6.0pt, yshift=0.2pt] at (0,1.3) {$Z^2$};
	
	\end{tikzpicture}
\end{equation} 
where $Z_{\vec{r}}^1$ and $X_{\vec{r}}^1$ ($Z_{\vec{r}}^2$ and $X^2_{\vec{r}}$) act on $\mathcal{H}_1$ ($\mathcal{H}_2$).

This is simply the two Wen plaquette operators acting on each layer of spins separately. This model is a simple realization of $\mathbb{Z}_2 \times \mathbb{Z}_2$ topological order. Anyons $e_1$ and $m_1$ correspond to the layer labeled by 1 in the model, $e_2$ and $m_2$ correspond to layer labeled by 2. 

We will consider that the lengths of the lattice have even size, so all the checkerboard pattern can be used here for both layers.

\subsection{Defects $\sigma$}

We begin with the simplest anyonic symmetry, $\sigma$, which is a direct generalization of the symmetry implemented in the toric code. In this case, we introduce a dislocation that breaks the translational symmetry of the lattice, in order to destroy the checkerboard pattern. The construction closely follows that of Sec.~\ref{sec3}. An array of spins is inserted along a branch cut, and the endpoints of this defect are associated with pentagon operators given by
\begin{equation}
	P^{\pm}_1 = \pm\begin{tikzpicture}[baseline={(current bounding box.center)}]

	\def\L{1.3}

	\coordinate (A) at (0,0);
	\coordinate (B) at (\L,0);
	\coordinate (C) at (\L,\L);
	\coordinate (D) at (0,\L);
	\coordinate (E) at (-0.5*\L,0.5*\L);

	\draw (A) -- (B) -- (C) -- (D) -- (E) -- cycle;

	\node[below left, xshift=6.0pt, yshift=-1.5pt] at (A) {$X^1_{\vec{r}}$};
	\node[below right, xshift=-6.0pt, yshift=-1.5pt] at (B) {$Z^1$};
	\node[above right, xshift=-6.0pt, yshift=1.5pt] at (C) {$X^1$};
	\node[above left, xshift=6.0pt, yshift=1.5pt] at (D) {$Z^1$};
	\node[left, xshift=-1.5pt] at (E) {$Y^1$};
	
	\end{tikzpicture},  \quad \quad \text{and} \quad \quad 	P^{\pm}_2 = \pm\begin{tikzpicture}[baseline={(current bounding box.center)}]

		\def\L{1.3}

		\coordinate (A) at (0,0);
		\coordinate (B) at (\L,0);
		\coordinate (C) at (\L,\L);
		\coordinate (D) at (0,\L);
		\coordinate (E) at (-0.5*\L,0.5*\L);

		\draw (A) -- (B) -- (C) -- (D) -- (E) -- cycle;

		\node[below left, xshift=6.0pt, yshift=-1.5pt] at (A) {$X^2_{\vec{r}}$};
		\node[below right, xshift=-6.0pt, yshift=-1.5pt] at (B) {$Z^2$};
		\node[above right, xshift=-6.0pt, yshift=1.5pt] at (C) {$X^2$};
		\node[above left, xshift=6.0pt, yshift=1.5pt] at (D) {$Z^2$};
		\node[left, xshift=-1.5pt] at (E) {$Y^2$};
		
		\end{tikzpicture}
\end{equation}

These pentagon operators commute with all other terms in the Hamiltonian. As discussed previously, the double loop operators around a single defect, $\hat{\Theta}_{e_1}$ and $\hat{\Theta}_{e_2}$, corresponding to the double loop of an $e_1$ and $e_2$ excitations, respectively, provide the appropriate quantum numbers to distinguish defect types (in appendix \ref{apC}, we provide an explicit calculation of the eigenvalues of the double loop operators algebraically). Here, the label $a$ corresponds to the species label $\lambda_a \in \frac{\mathcal{A}_1 \times \mathcal{A}_2}{ (1-\sigma)\mathcal{A}_1 \times \mathcal{A}_2}$. Thus, a pair of eigenvalues is required to distinguish the four defect species.

To distinguish bulk anyon charges in the parent state, we use the eigenvalues of the Wilson loop operators $\Gamma_{e_1}$, $\Gamma_{e_2}$, $\Gamma_{m_1}$, and $\Gamma_{m_2}$.

For the pair of pentagon operators $P_1^-$ and $P_2^-$, the eigenvalues of the double loop operators are $(\hat{\Theta}_{e_1}, \hat{\Theta}_{e_2}) = (-i, -i)$, corresponding to the defect $\sigma_0$. For $P_1^+$ and $P_2^-$, the eigenvalues are $(i, -i)$, corresponding to $\sigma_1$. For $P_1^-$ and $P_2^+$, they are $(-i, i)$, corresponding to $\sigma_2$. Finally, for $P_1^+$ and $P_2^+$, the eigenvalues are $(i, i)$, corresponding to $\sigma_3$.

The presence of dislocations modifies the ground state degeneracy, analogous to the discussion in Sec. \ref{secA}. Each pair of dislocations introduces two additional degrees of freedom that are not fixed by commuting stabilizers. Therefore, the ground state degeneracy changes by a factor of $2^2$. Consequently, each pair of defects contributes a quantum dimension $d_{\sigma_i} = 2$.

We now turn to the fusion rules. First, note that attaching a fermion $\psi_1$ or $\psi_2$ does not change the defect type. This is because the double loop operators either commute trivially or anticommute twice with the corresponding fermion string, yielding
\begin{equation}
	\label{5.4}
	\sigma_i \times \psi_1 = \sigma_i \times \psi_2 = \sigma_i.
\end{equation}

To compute $\sigma_i \times \sigma_i$, we use the identities involving products of double loop operators, as in Eq.~(\ref{3.6}). In the present case, these take the form
\begin{equation}
	\label{5.5}
\Gamma_{e_1}\Gamma_{m_1}=	-\Theta_{e_1}\Theta_{e_1}  = 1,
\end{equation}
and
\begin{equation}
\label{5.6}
	 \Gamma_{e_2}\Gamma_{m_2}=-\Theta_{e_2}\Theta_{e_2}  = 1.
\end{equation}
The first equation restricts the fusion channels to those carrying either trivial charge or the same charge as $\psi_1$, and similarly for the second equation with $\psi_2$. Therefore, the allowed fusion channels are $1$, $\psi_1$, $\psi_2$, and $\psi_1 \psi_2$. Consistency with Eq. (\ref{5.4}) leads to
\begin{equation}
	\sigma_i \times \sigma_i = 1+\psi_1+\psi_2+\psi_1\psi_2.
\end{equation}

This fusion rule is consistent with the previously obtained quantum dimension. Moreover, the allowed fusion channels coincide precisely with the anyons that are invariant under the anyonic symmetry $\sigma$.

\subsection{Defects $\rho$}

We now turn to the symmetry $\rho$, which corresponds to exchanging the two layers of the model. Here, the two layers will have a nontrivial interplay.

To implement this symmetry on the lattice, we introduce a branch cut along a line that terminates at the centers of two plaquettes. The endpoints of this line host the defects. Along the branch cut, the plaquette operators are modified as
\begin{equation}
	\tilde{F}^1_{\vec{r}} =  \begin{tikzpicture}[baseline={(current bounding box.center)}]

	\draw (0,0) -- (1.3,0) -- (1.3,1.3) -- (0,1.3) -- cycle;

	\node[below left, xshift=6.0pt, yshift=0.2pt] at (0,0) {$X^2_{\vec{r}}$};
	\node[below right, xshift=-6.0pt, yshift=-0.2pt] at (1.3,0) {$Z^2$};
	\node[above right, xshift=-6.0pt, yshift=0.2pt] at (1.3,1.3) {$X^1$};
	\node[above left, xshift=6.0pt, yshift=0.2pt] at (0,1.3) {$Z^1$};
	
	\end{tikzpicture}, \quad \quad \text{and} \quad \quad \tilde{F}^2_{\vec{r}} =  \begin{tikzpicture}[baseline={(current bounding box.center)}]

	\draw (0,0) -- (1.3,0) -- (1.3,1.3) -- (0,1.3) -- cycle;

	\node[below left, xshift=6.0pt, yshift=-0.2pt] at (0,0) {$X^1_{\vec{r}}$};
	\node[below right, xshift=-6.0pt, yshift=-0.2pt] at (1.3,0) {$Z^1$};
	\node[above right, xshift=-6.0pt, yshift=0.2pt] at (1.3,1.3) {$X^2$};
	\node[above left, xshift=6.0pt, yshift=0.2pt] at (0,1.3) {$Z^2$};
	
	\end{tikzpicture}
\end{equation}

\begin{figure}[h]
	\centering
	\includegraphics[width=0.63\linewidth]{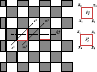}
	\caption{Defect line on the lattice that realizes the anyonic symmetry $e_1 \leftrightarrow e_2$ and $m_1 \leftrightarrow m_2$. The endpoints localize the modified operator $\tilde{Q}^{\pm}$, and the red squares that intersect the line are modified in a way that anyons crossing them change layers.}
	\label{fig:defect2}
\end{figure}

At the endpoints, the plaquette operators must be further modified to ensure commutation with the Hamiltonian. They are replaced by
\begin{equation}
	\tilde{Q}^{\pm} = \pm \begin{tikzpicture}[baseline={(current bounding box.center)}]

	\draw (0,0) -- (1.3,0) -- (1.3,1.3) -- (0,1.3) -- cycle;

	\node[below left, xshift=10.0pt, yshift=-0.2pt] at (0,0) {$X^1_{\vec{r}}X^2_{\vec{r}}$};
	\node[below right, xshift=-10.0pt, yshift=-0.2pt] at (1.3,0) {$Z^1Z^2$};
	\node[above right, xshift=-10.0pt, yshift=0.2pt] at (1.3,1.3) {$X^1X^2$};
	\node[above left, xshift=10.0pt, yshift=0.2pt] at (0,1.3) {$Z^1Z^2$};
	
	\end{tikzpicture}
\end{equation}
In addition, the plaquette operators adjacent to the endpoints may carry a positive or negative sign. The resulting branch cut and defect configuration are illustrated in Fig.~\ref{fig:defect2}.

The Hamiltonian remains a sum of commuting projectors, but the presence of the branch cut enforces that an anyon crossing it is transformed according to the symmetry $\rho$. In particular, an $e_1$ particle is mapped to $e_2$, and similarly for $m_1 \leftrightarrow m_2$. In this way, the branch cut implements a nontrivial coupling between the two layers.

The defect types are characterized by the double loop operators $\hat{\Theta}_{e_1}$ and $\hat{\Theta}_{m_1}$. One of these contracts to the local operator $\tilde{Q}^{\pm}$ at the defect, while the other reduces to a product of nearby plaquette operators adjacent to the endpoint. Their eigenvalues distinguish the defect species: $(-1,-1)$ that corresponds to $\rho_0$, $(1,-1)$ that corresponds to $\rho_1$, $(-1,1)$ that corresponds to $\rho_2$, and $(1,1)$, corresponding to $\rho_3$.

As in the previous case, the defects introduce additional ground state degeneracy. Each endpoint contributes a degree of freedom that is not fixed by an independent stabilizer. Although no new spins are added, one spin associated with the operator $\tilde{Q}^{\pm}_{\vec{r}}$ is not constrained by the set of stabilizer. This leads to a change of the ground state degeneracy by a factor of $4$, implying that each defect has quantum dimension $d_{\rho_i} = 2$.

Attaching the composite anyons $e_1 e_2$ or $m_1 m_2$ does not change the defect type. This is because the corresponding strings intersect the double loop operators twice, giving two anticommutations that cancel. Thus,
\begin{equation}
	\label{5.11}
	\rho_i \times e_1e_2 = \rho_i \times m_1m_2 = \rho_i.
\end{equation}

We now compute the fusion rules. As in the previous case, they follow from identities involving Wilson loop operators. However, in contrast to Eqs.~(\ref{5.5}) and (\ref{5.6}), there is no minus sign from self-intersections, since $e_1$ and $e_2$ (as well as $m_1$ and $m_2$) have trivial mutual braiding statistics. Therefore,
\begin{equation}
	\Gamma_{e_1}\Gamma_{e_2} = \Theta_{e_1} \Theta_{e_1} = 1
\end{equation}
\begin{equation}
	\Gamma_{m_1}\Gamma_{m_2} = \Theta_{m_1} \Theta_{m_1} = 1
\end{equation}

These constraints restrict the fusion channels to $1$, $e_1 e_2$, $m_1 m_2$, and $\psi_1 \psi_2$. By consistency with Eq. (\ref{5.11}), we obtain
\begin{equation}
	\rho_i \times \rho_i = 1+e_1e_2+m_1m_2+\psi_1\psi_2.
\end{equation}

Thus, the allowed fusion channels are precisely the anyons invariant under the action of the symmetry $\rho$.

\subsection{Defects $\xi$}

We now turn to the final anyonic symmetry, $\xi$, which combines layer exchange with electric-magnetic duality.

\begin{figure}
	\centering
	\includegraphics[width=0.6\linewidth]{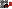}
	\caption{Branch cut that modifies the lattice by introducing extra degrees of freedom and a pentagon at the endpoints. The Hamiltonian is modified into one that is a sum of commuting projectors that allows the anyon to permute once they cross the cut line.}
	\label{fig:dislocation2}
\end{figure}

To implement this symmetry on the lattice, we again introduce a dislocation defect. Specifically, we insert a finite array of spins along a branch cut, whose endpoints host the defects, as illustrated in Fig. \ref{fig:dislocation2}.

The plaquette operators above the dislocation are defined as in the original model. However, below the defect line, they are modified to
\begin{equation}
		\tilde{F}_{\vec{r}}^1 = \begin{tikzpicture}[baseline={(current bounding box.center)}]

		\draw (0,0) -- (1.3,0) -- (1.3,1.3) -- (0,1.3) -- cycle;

		\node[below left, xshift=6.0pt, yshift=-0.2pt] at (0,0) {$X^2_{\vec{r}}$};
		\node[below right, xshift=-6.0pt, yshift=-0.2pt] at (1.3,0) {$Z^2$};
		\node[above right, xshift=-6.0pt, yshift=0.2pt] at (1.3,1.3) {$X^1$};
		\node[above left, xshift=6.0pt, yshift=0.2pt] at (0,1.3) {$Z^1$};
		
		\end{tikzpicture}, \quad \quad \text{and} \quad \quad  \tilde{F}^2_{\vec{r}} = \begin{tikzpicture}[baseline={(current bounding box.center)}]

		\draw (0,0) -- (1.3,0) -- (1.3,1.3) -- (0,1.3) -- cycle;

		\node[below left, xshift=6.0pt, yshift=-0.2pt] at (0,0) {$X^1_{\vec{r}}$};
		\node[below right, xshift=-6.0pt, yshift=-0.2pt] at (1.3,0) {$Z^1$};
		\node[above right, xshift=-6.0pt, yshift=0.2pt] at (1.3,1.3) {$X^2$};
		\node[above left, xshift=6.0-0.2pt, yshift=0.2pt] at (0,1.3) {$Z^2$};
		
		\end{tikzpicture}
\end{equation}

At the endpoints of the branch cut, the plaquette operators are replaced by pentagon operators
\begin{equation}
	P_1^{\pm} = \pm\begin{tikzpicture}[baseline={(current bounding box.center)}]

		\def\L{1.3}

		\coordinate (A) at (0,0);
		\coordinate (B) at (\L,0);
		\coordinate (C) at (\L,\L);
		\coordinate (D) at (0,\L);
		\coordinate (E) at (-0.5*\L,0.5*\L);

		\draw (A) -- (B) -- (C) -- (D) -- (E) -- cycle;

		\node[below left, xshift=6.0pt, yshift=-1.5pt] at (A) {$X_{\vec{r}}^1$};
		\node[below right, xshift=-6.0pt, yshift=-1.5pt] at (B) {$Z^1$};
		\node[above right, xshift=-6.0pt, yshift=1.5pt] at (C) {$X^1$};
		\node[above left, xshift=6.0pt, yshift=1.5pt] at (D) {$Z^1$};
		\node[left, xshift=-1.5pt] at (E) {$X^1 Z^2$};
		
	\end{tikzpicture},\quad\quad \text{and}\quad\quad P_2^{\pm}=\pm\begin{tikzpicture}[baseline={(current bounding box.center)}]

		\def\L{1.3}
		
		\coordinate (A) at (0,0);
		\coordinate (B) at (\L,0);
		\coordinate (C) at (\L,\L);
		\coordinate (D) at (0,\L);
		\coordinate (E) at (-0.5*\L,0.5*\L);

		\draw (A) -- (B) -- (C) -- (D) -- (E) -- cycle;

		\node[below left, xshift=6.0pt, yshift=-1.5pt] at (A) {$X_{\vec{r}}^2$};
		\node[below right, xshift=-6.0pt, yshift=-1.5pt] at (B) {$Z_2$};
		\node[above right, xshift=-6.0pt, yshift=1.5pt] at (C) {$X_2$};
		\node[above left, xshift=6.0pt, yshift=1.5pt] at (D) {$Z_2$};
		\node[left, xshift=-1.5pt] at (E) {$X^2Z^1$};
		
	\end{tikzpicture}
\end{equation}

When an anyon crosses the branch cut, its type is transformed according to the symmetry $\xi$, namely $e_1 \rightarrow m_2$ and $e_2 \rightarrow m_1$.

In this case the appropriate quantum numbers that characterize the defects are given by the double loop operators $\hat{\Theta}_{e_1}$ and $\hat{\Theta}_{e_2}$. Their pair of eigenvalues are sufficient to distinguish the four types of defects. The pair of operators $P_1^-$ and $P_2^-$ leads to $(\Theta_{e_1},\Theta_{e_2}) = (-1,-1)$, and we have the bare defect $\xi_0$, $P_1^+$ and $P_2^-$ give the eigenvalues $(1,-1)$ to the double loop operators, and correspond to the defect $\xi_1$, $P_1^-$ and $P_2^+$ give the eigenvalues $(-1,1)$ that corresponds to the defect $\xi_2$ and lastly, $P_1^+$ and $P_2^+$ give the eigenvalues $(1,1)$ that corresponds to the defect $\xi_3$. 

Attaching the anyons $e_1 m_2$ or $e_2 m_1$ does not change the defect type, since the corresponding strings intersect the double loops twice, producing two cancelling anticommutations. Thus,
\begin{equation}
	\label{5.19}
	\xi_i \times e_1m_2 = \xi_i \times m_2m_1 = \xi_i.
\end{equation}

As before, these defects have quantum dimension $d_{\xi_i} = 2$. Each pair of defects introduces two degrees of freedom that are not fixed by stabilizers, leading to a change by a factor of $4$ in the ground state degeneracy.

The fusion rules between defects follow by deforming the double loop operators around each defect to a pair of loops $\Gamma_{e_1}\Gamma_{m_2}$ and $\Gamma_{e_2}\Gamma_{m_1}$. Since the symmetric anyons under $\xi$ have trivial braiding, there is no phase accumulated in this deformation. So, we have
\begin{equation}
	\Gamma_{e_1}\Gamma_{m_2} = \Theta_{e_1} \Theta_{e_1} = 1
\end{equation}
\begin{equation}
	\Gamma_{e_2}\Gamma_{m_1} = \Theta_{e_2} \Theta_{e_2} = 1
\end{equation}

These constraints lead to four possible fusion channels $1$, $e_1m_2$, $e_2m_1$, and $\psi_1\psi_2$. From Eq. (\ref{5.19}), we have
\begin{equation}
	\xi_i \times \xi_i = 1+ e_1m_2+e_2m_1+\psi_1\psi_2.
\end{equation}
In all cases, defect fusion closes onto the invariant anyons under the corresponding anyonic symmetry.

\section{Fusion Rules Between $\sigma$ and $\rho$}

In this section, we derive the fusion rules between defects associated with different anyonic symmetries. In particular, we focus on the electric-magnetic symmetry $\sigma$ and the layer exchange symmetry $\rho$. It suffices to consider the fusion of the corresponding bare defects, since all other fusion channels can be obtained by attaching Abelian anyons. Thus, we analyze the fusion $\sigma_0 \times \rho_0$.

First, observe that when the two defects are brought close together, their combined action corresponds to the composition of the two symmetries, $\xi = \sigma \circ \rho$. Indeed, an anyon crossing both defects undergoes the sequence of transformations $e_{1/2} \rightarrow m_{1/2} \rightarrow m_{2/1}$, which is precisely the action of $\xi$. Therefore, the fusion channels must lie in the $\xi$-defect sector, so they are of the form $\xi_i$.

Since each defect has quantum dimension $d = 2$, the total quantum dimension of the fusion product is $d_{\sigma_0} d_{\rho_0} = 4$. As each $\xi_i$ has quantum dimension 2, we expect two fusion channels.

To determine which channels appear, we evaluate the double loop operators corresponding to $e_1$ and $e_2$, which characterize the $\xi$ defects. These loops encircle both defects, and their eigenvalues can be computed using the plaquette and pentagon operators. One finds
\begin{equation}
	\Theta_{e_1}\Theta_{e_2} = i^2 = -1.
\end{equation}
The $i^2$ comes from the identities $Z^iX^i = iY^i$. The only possible solutions for $(\Theta_{e_1},\Theta_{e_2})$ are $(1,-1)$ and $(-1,1)$. They correspond to the defects $\xi_1$ and $\xi_2$, respectively. Note that the fusion outcome cannot consist of a single channel with multiplicity 2, say $\sigma_0\times\rho_0 =2 \xi_1$, since, from Eq. (\ref{5.11}) fusion with $e_1e_2$ would lead to $\sigma_0 \times\rho_0 = 2 \xi_2$, which is inconsistent with the original fusion rule. Therefore, the fusion rule is given by
\begin{equation}
	\label{6.2}
	\sigma_0 \times \rho_0 = \xi_1+\xi_2.
\end{equation}

This result admits a natural interpretation, the fusion $\sigma \times \rho \rightarrow \xi$ involves a nontrivial trivalent junction that can carry an Abelian anyon. The appearance of these fusion channels reflects the fact that the symmetries are realized projectively, leading to an Abelian anyon trapped at the junction of the fusion. A more detailed discussion of this structure is provided in Appendix~\ref{apB}.

\section{Conclusions}

In this work, we addressed the problem of realizing twist defects in a $\mathbb{Z}_2 \times \mathbb{Z}_2$ topological order within a microscopic system. This phase has a rich anyonic structure, which leads to an extensive set of symmetries. We constructed lattice dislocations associated with three particularly interesting types of anyonic symmetries, which are the electric-magnetic duality, layer exchange, and their composition. On a double Wen plaquette model, dislocations were introduced by considering branch cuts that modify the Hamiltonian into one that keeps its exact solvability, while allowing for localized defects.  

All  three anyonic symmetries are $\mathbb{Z}_2$, twofold, symmetries and they lead to the same defect quantum dimensions, equal to two. To characterize the non-Abelian static modes, we use double loop operators that are defined by the worldlines of anyons that need to wind the defect twice in order to close into a symmetry operator.

Relatedly, the color code also realizes a $\mathbb{Z}_2\times\mathbb{Z}_2$ topological order and its twist defects have been studied in \cite{Kesselring:2018jbv}. There, it was shown that when the twist is capable of localizing four anyons,\footnote{An anyon can be localized by a twist defect if it is invariant under the symmetry.} it has quantum dimension equal to two, which agrees with our results.

Fusion rules between defects realizing different types of anyonic symmetries were also  computed by using loop operators. In the cases considered in this work, the fusion rule between a defect endowed with electric-magnetic duality and one endowed with layer exchange symmetry results in two fusion channels of defects endowed with the composition of these symmetries. We also find that these fusion processes act projectively due to an Abelian anyon trapped at the trivalent junction of the fusion.

The simultaneous realization of the three twist defect types may be relevant to defect based topological quantum computation. Although the intrinsic anyons of the $\mathbb{Z}_2\times\mathbb{Z}_2$ topological order are Abelian, twist defects can support non-Abelian fusion spaces, allowing quantum information to be encoded nonlocally in the fusion channels of several defects. Braiding these defects implements unitary operations on the degenerate fusion space, providing the basic ingredient for topological quantum gates. The coexistence of several defect types also permits mixed fusion processes and suggests a richer class of braiding operations that are absent in constructions supporting only a single twist defect.

\begin{acknowledgments}
The author would like to thank Pedro Gomes for very useful discussions. The author also acknowledges the financial support from the Brazilian funding agency CAPES.

\end{acknowledgments}

\appendix

\section{Defect Types from Quotient Groups \label{apA}}

In this appendix, we explain how to obtain the defect types associated with the relevant subgroup of anyonic symmetries studied in this work by  using a quotient group construction. This method was used in Eqs. \ref{4.9}, \ref{4.10}, and \ref{4.11}. For an Abelian topological phase with anyon content $\mathcal{A}$ forming a group under fusion, this procedure systematically determines the defect labels.

To illustrate the method, we explicitly construct the defects associated with the anyonic symmetry $\sigma$, which corresponds to the electric-magnetic duality. The defect types are labeled by elements of the quotient group
\begin{equation}
	\frac{\mathcal{A}}{(1-\sigma)\mathcal{A}}
\end{equation}
This quotient consists of cosets of the subgroup $(1-\sigma)\mathcal{A}$, which we now determine.

Since $\sigma^{-1} = \sigma$, the subgroup $(1-\sigma)\mathcal{A}$ is generated by elements of the form $a \times \sigma(a)$. Applying this to the generators $\langle e_1, e_2, m_1, m_2 \rangle$, we find
\begin{equation}
	(1-\sigma)\langle e_1,e_2,m_1,m_2 \rangle = \langle \psi_1,\psi_2 \rangle = \{1,\psi_1,\psi_2,\psi_1\psi_2\}.
\end{equation}

Since this subgroup has 4 elements and the full anyon group $\mathcal{A}$ has 16 elements, the quotient group has 4 elements, corresponding to 4 distinct defect types. These are represented by the cosets
\begin{eqnarray}
	&&\overline{1} = 1\times (1-\sigma)\mathcal{A} = \{1,\psi_1,\psi_2,\psi_1\psi_2\}\\
	&&\overline{e_1 } = e_1 \times (1-\sigma)\mathcal{A}  = \{e_1,m_1,e_1\psi_2,m_1\psi_2\} \\
	&&\overline{e_2} = e_2 \times (1-\sigma)\mathcal{A} = \{e_2, e_2 \psi_1, m_2, m_2\psi_1\}\\
	&&\overline{e_1e_2} = e_1e_2 \times (1-\sigma)\mathcal{A} =\{e_1e_2,e_2m_1,e_1m_2,m_1m_2\}
\end{eqnarray}
Thus, the defect labels associated with $\sigma$ are
\begin{equation}
	\frac{\mathcal{A}}{(1-\sigma)\mathcal{A}} = \{\overline{1},\overline{e_1 } ,\overline{e_2},\overline{e_1e_2}\}.
\end{equation}

For the anyonic symmetry $\rho$, the procedure is analogous. The quotient group is
\begin{equation}
	\frac{\mathcal{A}}{(1-\rho)\mathcal{A}} = \{\overline{1},\overline{e_1},\overline{ m_1},\overline{ \psi_1}\},
\end{equation} 
where the cosets are
\begin{eqnarray}
	&&\overline{1} = 1\times (1-\rho)\mathcal{A} = \{1,e_1e_2,m_1m_2,\psi_1\psi_2\}\\
	&&\overline{e_1} = e_1 \times (1-\rho)\mathcal{A}  = \{e_1,e_2,\psi_1m_2,m_1\psi_2\} \\
	&&\overline{ m_1} = m_1 \times (1-\rho)\mathcal{A} = \{m_1, \psi_1e_2, m_2, e_1\psi_2\}\\
	&&\overline{ \psi_1} = \psi_1 \times (1-\rho)\mathcal{A} =\{\psi_1,m_1e_2,e_1m_2,\psi_2\}.
\end{eqnarray}

Finally, for the symmetry $\xi$, the defect labels are obtained from
\begin{equation}
	\frac{\mathcal{A}}{(1-\xi)\mathcal{A}} = \{\overline{1},\overline{e_1} ,\overline{m_1},\overline{\psi_1}\},
\end{equation}
where the subgroup $(1-\xi)\mathcal{A}$ is given by $\{1, e_1 m_2, e_2 m_1, \psi_1 \psi_2\}$ and the corresponding cosets are
\begin{eqnarray}
	&&\overline{1} = 1\times (1-\xi)\mathcal{A} = \{1,e_1m_2,e_2m_1,\psi_1\psi_2\}\\
	&&\overline{e_1} = e_1 \times (1-\xi)\mathcal{A}  = \{e_1,m_2,\psi_1e_2,m_1\psi_2\} \\
	&&\overline{m_1} = m_1 \times (1-\xi)\mathcal{A} = \{m_1, \psi_1m_2,e_2, e_1\psi_2\}\\
	&&\overline{\psi_1} = \psi_1 \times (1-\xi)\mathcal{A} =\{\psi_1,m_1m_2,e_1e_2,\psi_2\}.
\end{eqnarray}
This completes the classification of the twist defect types associated with the three anyonic symmetries considered in this work.

\section{Defects Fusion Rules from Quotient Groups \label{apB}}

In this appendix, we show how the quotient group construction can also be used to determine fusion rules between defects associated with different anyonic symmetries. When defects corresponding to two distinct symmetries are fused, the resulting channels must belong to the defect sector of the composed symmetry.

As an example, consider the fusion of the bare defects $\sigma_0$ and $\rho_0$. (Fusion rules involving other defect types can be obtained by attaching Abelian anyons.) Their fusion can be expressed as
\begin{equation}
	\label{B1}
	\sigma_0 \times \rho_0 =e^{\sigma,\rho} \times \sum_{i \in \mathcal{A}_{\xi}^{\sigma,\rho}} \xi_{i},
\end{equation}
where $e^{\sigma,\rho}$ is an Abelian anyon localized at the trivalent junction formed by the intersection of the $\sigma$, $\rho$, and $\xi$ defect lines. This anyon encodes a possible projective phase arising from the composition of the symmetries\footnote{The action of the anyonic symmetry $\xi$ on the superselection sector of an anyon $c$ can be represented by an operator $\hat{\xi}$. When the symmetry is decomposed as $\xi = \sigma \rho$, the operators need not compose linearly. Instead they may furnish a projective phase, $\hat{\xi} = e^{i\phi_{\sigma,\rho}(c)}\hat{\sigma}\hat{\rho}$, that can be interpreted as the braiding phase between $c$ and an Abelian anyon $e^{\sigma,\rho}$.} \cite{Teo:2015xla,Barkeshli:2014cna}. The sum runs over the quotient set
\begin{equation}
	\label{B2}
	\mathcal{A}_{\xi}^{\sigma,\rho} = \frac{(1-\sigma)\mathcal{A}+(1-\rho)\mathcal{A}}{(1-\xi)\mathcal{A}},
\end{equation}
Here, the numerator $(1-\sigma)\mathcal{A} + (1-\rho)\mathcal{A}$ denotes the subgroup generated by combining elements of the two subgroups $(1-\sigma)\mathcal{A}$ and $(1-\rho)\mathcal{A}$. Explicitly, this subgroup is
\begin{equation}
	(1-\sigma)\mathcal{A}+(1-\rho)\mathcal{A} = \{1,\psi_1,\psi_2,e_1e_2,m_1m_2,\psi_1\psi_2,e_2m_1,e_1m_2\}.
\end{equation}
Since this subgroup has 8 elements and $(1-\xi)\mathcal{A}$ has 4 elements, the quotient in Eq.~\ref{B2} contains 2 elements. These correspond to the cosets
\begin{eqnarray}
	&&\overline{1} = 1 \times (1-\xi)\mathcal{A}= \{1,e_1m_2,e_2m_1,\psi_1\psi_2\}\\
	&&\overline{\psi_1} = \psi_1 \times (1-\xi)\mathcal{A} = \{\psi_1,m_1m_2,e_1e_2,\psi_2\}.
\end{eqnarray}

Therefore, in the absence of a projective phase, $e^{\sigma,\rho} = 1$, Eq.~\ref{B1} gives
\begin{equation}
	\sigma_0 \times \rho_0 = \xi_0 + \xi_3.
\end{equation}

\begin{figure}
	\centering
	\includegraphics[width=0.4\linewidth]{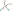}
	\caption{Fusion between defects endowed with symmetries $\sigma$ and $\rho$, resulting in channels of defects endowed with the symmetry $\xi$. The trivalent junction carries an Abelian anyon $a$.}
	\label{fig:trivalentjunction}
\end{figure}

However, if a nontrivial projective phase is present, an Abelian anyon is effectively bound to the trivalent junction, as depicted in Fig. \ref{fig:trivalentjunction}. This shifts the fusion outcomes by that anyon. Thus, a projective phase can lead to the fusion rule
\begin{equation}
	\sigma_0 \times \rho_0 = \xi_1 + \xi_2,
\end{equation}
which agrees with the result obtained in Eq.~\ref{6.2}.

\section{\label{apC}Double Loop Around Defects: Calculation}

In this appendix, we show how to compute the phase acquired by a double loop operator when an anyon encircles a defect. These phases play a crucial role in characterizing the different defect types. We illustrate the procedure for the case of the electric-magnetic duality symmetry $\sigma$. In the vector representation introduced in Sec. \ref{secIII}, this symmetry is represented by the matrix
\begin{equation}
\label{matrix}
	\begin{pmatrix}
		\sigma^x & 0 \\
		0 & \sigma^x
	\end{pmatrix},
\end{equation}
since it exchanges the first and second components of the vector, corresponding to $e_1$ and $m_1$, respectively, as well as the third and fourth components, corresponding to $e_2$ and $m_2$, respectively.

Consider an Abelian anyon $a$ ($a$ may also denote the vector representation of the anyon) encircling a defect $\sigma_\lambda$ labeled by a species vector $t^\lambda$, which encodes the charge attached to the defect. The corresponding double loop operator is denoted by $\hat{\Theta}_a^\lambda$. By absorbing the string attached to the defect into the definition of the operator, this can be written as
\begin{equation}
	\label{C2}
	\hat{\Theta}_a^{\lambda} = e^{2 \pi i a^TK^{-1}t^\lambda} \hat{\Theta}_a^0,
\end{equation}
where $\hat{\Theta}_a^0$ is the double loop operator around a bare defect, and $K$ is the matrix of the $K$-matrix Chern-Simons that describe two copies of the effective theories of toric code, or equivalently, two BF actions. It is equal to the matrix in Eq. (\ref{matrix}) up to a factor of two, and it is the appropriate choice to describe the $\mathbb{Z}_2 \times \mathbb{Z}_2$ topological order considered in this work within an effective topological field theory \cite{PhysRevB.46.2290}.

To compute $\hat{\Theta}_a^0$, we use three identities that fix its eigenvalues. First, invariance under the twofold symmetry implies  $\hat{\Theta}_{\sigma a}^0 = \hat{\Theta}_a^0$.
Second, the auto-intersection identity gives
\begin{equation}
	\hat{\Theta}_{a+b}^0= e^{2 \pi i a^TK^{-1}(\sigma b)} \hat{\Theta}_a\hat{\Theta}_b,
\end{equation}
Third, the self-linking relation reads
\begin{equation}
	\label{c4}
	\left(\hat{\Theta}_a^0 \right)^2 = e^{2 \pi i a^TK^{-1}(\sigma a)}.
\end{equation}
The auto intersection identity suggests a solution of the form
\begin{equation}
	\Theta_a^0 = e^{\pi ia^T K^{-1}(\sigma a) +2 \pi i a^T K^{-1} \mathcal{V}}.
\end{equation}
So the problem reduces to finding the vector $\mathcal{V}$. Imposing invariance under two-fold symmetry leads to $\mathcal{V} = \sigma \mathcal{V}$. So the vector $\mathcal{V}$ has the form
\begin{equation}
	\mathcal{V} = \alpha \begin{pmatrix}
		1 \\
		1\\
		0\\
		0
	\end{pmatrix} + \beta \begin{pmatrix}
	0\\
	0\\
	1\\
	1
	\end{pmatrix} = \alpha t_1 + \beta t_2.
\end{equation}
Now, Eq. \ref{c4} leads to
\begin{equation}
	e^{4\pi i a^TK^{-1}(\alpha t_1+\beta t_2)} = e^{2\pi i a^T(\alpha t_1 +\beta t_2)}= 1,
\end{equation}
which must hold for all anyons $a$.
The minimal nontrivial solution allowed by this condition is $\alpha = 1$ and $\beta=1$, so the eigenvalue of the double loop operator is given by
\begin{equation}
	\label{C8}
	\Theta_a^0 = e^{2\pi i a^TK^{-1}\sigma(\frac{1}{2}a+t_1+t_2)}.
\end{equation}

Finally, the eigenvalues of the double loop operators around a general defect $\sigma_\lambda$ follow directly from Eqs.~\ref{C2} and \ref{C8}. These phases fully characterize the different electric-magnetic twist defect types.

\bibliographystyle{unsrt}
\bibliography{ref}

\end{document}